\documentclass[journal,twoside,web]{ieeecolor}
\usepackage{generic}
\usepackage{cite}
\usepackage{amsmath,amssymb,amsfonts}
\usepackage{algorithmic}
\usepackage{graphicx}
\usepackage{algorithm,algorithmic}
\usepackage{bbding}
\makeatletter
\let\NAT@parse\undefined
\makeatother
\usepackage[
hidelinks,
colorlinks = true,
linkcolor = blue,
citecolor = blue,
urlcolor = black
]{hyperref}

\usepackage{textcomp}
\usepackage{array}
\usepackage{booktabs}
\usepackage{multirow}
\def\BibTeX{{\rm B\kern-.05em{\sc i\kern-.025em b}\kern-.08em
    T\kern-.1667em\lower.7ex\hbox{E}\kern-.125emX}}
\begin{document}
\title{TFCDiff: Robust ECG Denoising via Time-Frequency Complementary Diffusion}
\author{Pengxin Li, Yimin Zhou, Jie Min, Yirong Wang, Wei Liang, Qingling Xia and Wang Li
\thanks{This work was supported in part by the Chongqing Municipal Key Project for Technology Innovation and Application Development under Grant CSTB2024TIAD-KPX0042. \textit{(Corresponding author: Wang Li)}}
\thanks{Pengxin Li, Yimin Zhou, Jie Min, Yirong Wang, Wei Liang, and Wang Li are with the School of Pharmacy and Bioengineering, Chongqing University of Technology, Chongqing 400054, China (e-mail: wang.l@cqut.edu.cn).}
\thanks{Qingling Xia is with the School of Artificial Intelligence, Chongqing University of Technology, Chongqing 401135, China (e-mail: qingling@cqut.edu.cn).}}

\maketitle

\begin{abstract}
Ambulatory electrocardiogram (ECG) readings are prone to mixed noise from physical activities, including baseline wander (BW), muscle artifact (MA), and electrode motion artifact (EM). Developing a method to remove such complex noise and reconstruct high-fidelity signals is clinically valuable for diagnostic accuracy. However, denoising of multi-beat ECG segments remains understudied and poses technical challenges. To address this, we propose Time-Frequency Complementary Diffusion (TFCDiff), a novel approach that models ECG signals in the Discrete Cosine Transform (DCT) domain. To refine waveform details, we incorporate Temporal Feature Enhancement Mechanism (TFEM) to reinforce temporal representations and preserve key physiological information. Comparative experiments on a synthesized dataset demonstrate that TFCDiff achieves state-of-the-art performance across five evaluation metrics. Furthermore, TFCDiff shows superior generalization on the unseen SimEMG Database, outperforming all benchmark models. Notably, TFCDiff processes raw 10-second sequences and maintains robustness under flexible random mixed noise (fRMN), enabling plug-and-play deployment in wearable ECG monitors for high-motion scenarios. Source code is available at https://github.com/Miroircivil/TFCDiff.
\end{abstract}

\begin{IEEEkeywords}
ECG denoising, conditional diffusion model, discrete cosine transform, spectral modeling, random mixed noise.
\end{IEEEkeywords}

\section{Introduction}
\label{sec:introduction}
\IEEEPARstart{C}{ardiovascular} diseases (CVDs) remain a critical public health challenge, affecting over 535 million people and causing 19.8 million deaths globally in 2022 alone \cite{bib1}. The electrocardiogram (ECG), an efficient and non-invasive tool for recording cardiac electrical activity, aids physicians in detecting conditions such as atrial fibrillation, myocardial ischemia, and myocardial infarction \cite{bib2}. Early detection enables timely interventions and appropriate therapies for CVDs. Recently, wearable devices have expanded the applications of ECG monitoring, extending into sports cardiology through fitness trackers, chest straps and smartwatches \cite{bib3,bib4}. The real-time data generated by these platforms offer valuable clinical insights for personalized treatment \cite{bib5}.

However, during physical activity, ECG readings are prone to mixed noise, including baseline wander (BW), muscle artifact (MA), and electrode motion artifact (EM), which exhibit significantly higher amplitudes than those observed at rest \cite{bib6}. These noises can significantly degrade ECG signal quality, impairing diagnostic accuracy. Various time-frequency signal processing methods have been employed to address these noise issues \cite{bib7}, \cite{bib8}, \cite{bib9}. Unfortunately, they perform poorly under high-intensity noise due to the limited separability between noise and ECG components.

In recent studies, data-driven deep learning methods have been increasingly applied to ECG denoising. Antczak proposed an LSTM-based Deep Recurrent Neural Network (DRNN) for end-to-end ECG denoising and demonstrated superior performance over traditional digital filters \cite{bib10}. Inspired by the Inception module, Romero et al. introduced DeepFilter that incorporates a Multi-Kernel Linear and Non-Linear (MKLANL) filter module to handle noisy ECG signals with multi-scale features \cite{bib11}. Hu et al. proposed a lightweight U-Net (LUNet) to remove noise in ECG signals \cite{bib12}. Denoising Autoencoders (DAEs) have also been notably applied, with variants such as Fully Convolutional Network (FCN-DAE) \cite{bib13}, Convolutional Block Attention Module (CBAM-DAE) \cite{bib14}, and Attention-based Convolutional (ACDAE) \cite{bib15}. Chen et al. subsequently proposed a Transformer-based Convolutional DAE (TCDAE), achieving remarkable results in ECG noise removal \cite{bib16}. Generative models capable of learning the underlying data distribution have also been utilized for ECG signal restoration. For instance, Wang et al. proposed a conditional Generative Adversarial Network (CGAN) that eliminates noise while preserving the morphological fidelity of ECG signals \cite{bib17}. Additionally, a series of diffusion-based methods, including Deep Score-Based Diffusion Model (DesCod) \cite{bib18}, ECG Denoising Diffusion Model (EDDM) \cite{bib19}, and Improved Denoising Diffusion Probabilistic Model (IDPM) \cite{bib20}, have emerged.

\begin{figure*}[!t]
\centerline{\includegraphics[width=0.8\textwidth]{figures/Diffusion_scheme.pdf}}
\caption{Schematic of the TFCDiff workflow. During training, the clean DCT coefficients $\boldsymbol{d}_0$ is corrupted by the forward diffusion process $q$, and the noise predictor, conditioned on the noisy observation $\tilde{\boldsymbol{d}}$, learns to predict the added noise. During sampling, a random Gaussian noise $\boldsymbol{d}_T$ is iteratively denoised via the reverse process $p$ to reconstruct the denoised coefficients.}
\label{fig1}
\end{figure*}

Due to the lack of evaluation datasets, deep learning methods typically rely on synthesized signals, which are generated by adding pure noise to clean signals \cite{bib21}. Different methods vary in their choice of noise types. For instance, DeepFilter, DesCod, and IDPM only used BW. EDDM, CGAN, and LUNet incorporated three types of noise (BW, MA and EM) but trained and tested on each separately, FCN-DAE and CBAM-DAE combined these noise types with equal weights. TCDAE introduced a random mixed noise (RMN) strategy, which randomly selected one or more noise types and combined them with uniform weights to mimic real-world noise conditions.

Despite numerous algorithms, they are plagued by various limitations. Many models are designed to use single-beat sequences, a methodology that impedes practical implementation \cite{bib10}, \cite{bib11}, \cite{bib14}, \cite{bib17}, \cite{bib18}, \cite{bib19}. Most deep learning methods are trained on synthesized datasets that include only a single noise type or a uniform mixture of noise types, failing to account for varied combination ratios and thus lacking flexibility in simulating diverse noise distributions. ECG signals are characterized by their unique periodic waveforms, such as P-QRS-T complexes, yet current deep learning models generally lack specialized architectures for the frequency domain. TCDAE incorporated a frequency-weighted Huber loss function, but still lacked a spectrum-specific architecture in the neural network. These limitations hinder the accurate reconstruction of noisy ECG signals in real-world applications. In this paper, we propose Time-Frequency Complementary Diffusion (TFCDiff) for multi-beat ECG denoising. TFCDiff applies the discrete cosine transform (DCT) to ECG signals and constructs a conditional diffusion model in the frequency domain. Modeling in the frequency domain enables TFCDiff to precisely restore ECG waveform details under varying noise intensities while reducing computational overhead by truncating irrelevant high-frequency components. To better mimic real-world noise, we adopt a flexible random mixed noise (fRMN) strategy in our synthesized dataset, which assigns random weights to different noise types to enhance diversity. In comparative experiments with benchmark methods, TFCDiff achieves state-of-the-art (SOTA) performance in ECG denoising on both synthesized and real datasets. The main contributions of this work can be summarized into three aspects:
\begin{enumerate}
    \item We propose TFCDiff, a conditional diffusion model defined in the frequency domain that trains and performs inference on raw 10-second multi-beat sequences. 
    \item To enhance the quality of reconstructed signals, we introduce a one-dimensional (1D) U-Net incorporating Temporal Feature Enhancement Mechanism (TFEM) as the noise predictor. TFEM comprises Temporal Feature Extraction (TFE) and Temporal Feature Fusion (TFF) modules that refine waveform details by reinforcing temporal representations.
    \item Experimental results demonstrate superior generalization of TFCDiff, with the model achieving robust denoising performance on the real-world ECG noise dataset that is never used during training.
\end{enumerate}

\section{Related Works}
\label{sec:Related Works}
\subsection{Diffusion Model}
\label{subsec:Diffusion Model}
Diffusion models have emerged as a powerful class of deep generative models. The fundamental principle was introduced by \cite{bib22}, and the architecture was subsequently improved by \cite{bib23} and \cite{bib24}, achieving unprecedented success in image generation. Further developments by \cite{bib25} extended diffusion models to conditional generation, enhancing their applicability. To address the issue of slow generation, \cite{bib26} reformulated the sampling process as a deterministic ordinary differential equation (ODE), reducing sampling steps from thousands \cite{bib23} to just 10 -- 20. Additionally, \cite{bib27} introduced flow matching for building arbitrary distributions. \cite{bib28} and \cite{bib29} replaced the noise predictor in U-Net with a transformer, offering greater modeling capacity. Notably, \cite{bib30} pioneered the application of diffusion models in the frequency domain, orthogonal to previous studies.

Recently, \cite{bib18} initially applied conditional diffusion models to ECG noise removal. Building on this, \cite{bib19} introduced a dual-path diffusion process that separates ECG noise diffusion from Gaussian noise diffusion, while \cite{bib20} optimized the noise predictor through pruning techniques. In this paper, we adopt the diffusion model in the frequency domain due to its superior performance in restoring the characteristic waveform of ECG.

\subsection{Modeling in Frequency Domain}
Neural networks often utilize spectral modeling as a module to accelerate computation and enrich feature representation. For instance, Tatsunami et al. substituted Multi-head Self-attention (MHSA) with a dynamic filter based on Fourier Transform for efficient high-resolution image recognition \cite{bib31}. In denoising tasks, Kong et al. achieved high-quality image deblurring through frequency domain transformers \cite{bib32}, while \cite{bib33} reconstructed high-fidelity RGB images by fusing spectral features. Beyond utilizing frequency modules, some studies directly modeled source data in the frequency domain to learn its distribution. The JPEG-LM proposed by \cite{bib34} significantly enhances image generation by training Large Language Model (LLM) with JPEG codec-encoded images. In \cite{bib30}, a diffusion model established in the DCT domain enabled low-cost and high-quality image generation. Buchholz et al. applied an autoregressive model in the Fourier domain to tackle image super-resolution tasks \cite{bib35}. Building upon the aforementioned approaches, we develop a diffusion model defined in the DCT domain and introduce TFEM to refine ECG waveform details.

\section{Methods}
Fig.\autoref{fig1} presents an overview of the proposed TFCDiff workflow. Our methodology comprises a training phase and a sampling phase, and the entire process is conducted in the frequency domain. In the training phase, the objective is to optimize a U-Net-based noise predictor that estimates the Gaussian noise added to the clean DCT coefficients, using a corresponding noisy observation as the conditioning input. In the sampling phase, we begin with random Gaussian noise and iteratively reconstruct it using the trained noise predictor, guided by the DCT coefficients of the noisy ECG signal to be denoised. To enhance the noise predictor, we incorporate TFEM that consists of TFE and TFF modules applied in a stream of spectral feature maps.

\subsection{DCT and Truncating}
\label{subsec:DCT and Truncating}
The Discrete Cosine Transform (DCT) decomposes a real signal into real-valued frequency coefficients by extending it to an even function \cite{bib36}. Due to its superior energy compaction and computational efficiency, DCT has become a cornerstone of image compression standards, most notably JPEG \cite{bib37}. A detailed comparison of DCT against other time-frequency transformation methods such as Fast Fourier Transform (FFT) and Discrete Wavelet Transform (DWT) is provided in Supplementary Note 1.

\begin{figure}[!t]
\centerline{\includegraphics[width=\columnwidth]{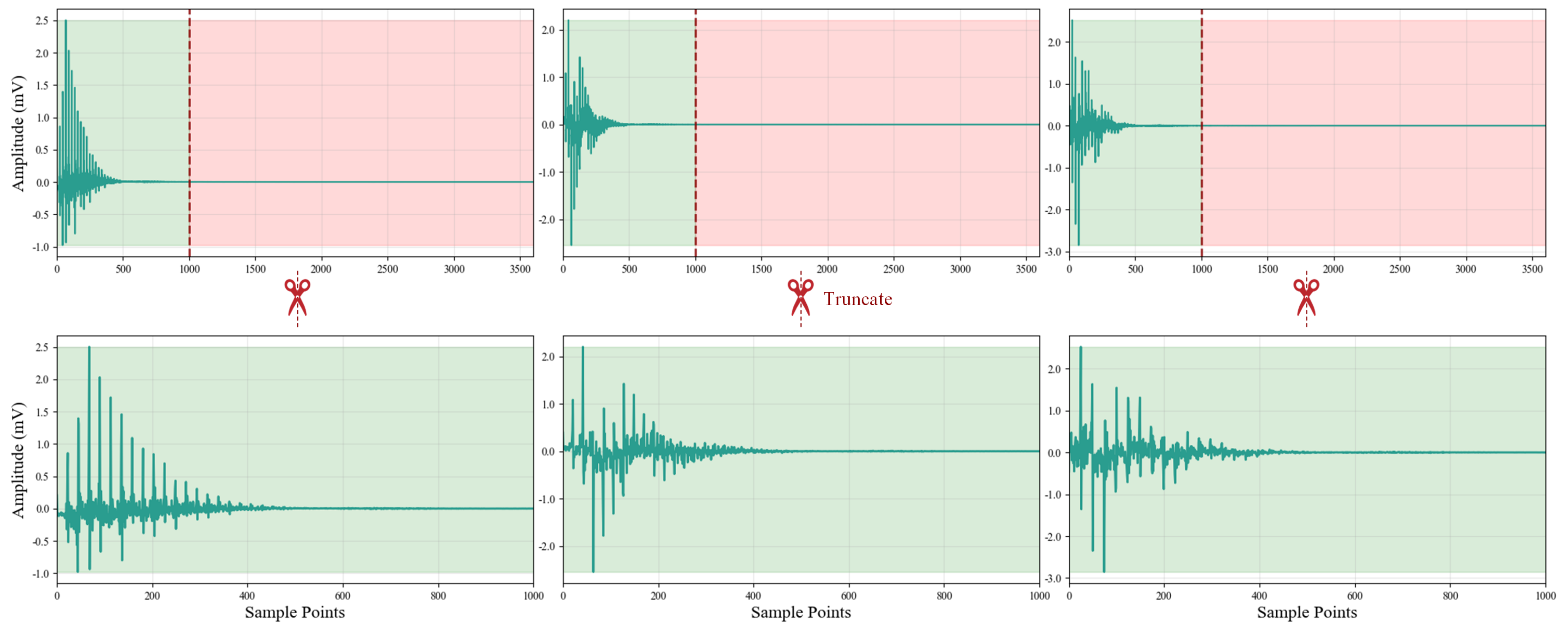}}
\caption{Truncation of DCT coefficients for 10-s signals sampled at 360 Hz by retaining the first 1000 coefficients to preserve frequency content below 50 Hz.}
\label{fig2}
\end{figure}

Given an ECG signal $\boldsymbol{l}\in {{\mathbb{R}}^{N}}$, we apply the 1D type-II DCT to obtain its frequency-domain representation $\boldsymbol{l}\in {{\mathbb{R}}^{N}}$. The original time-domain signal is recovered using the inverse DCT (IDCT), implemented via the type-III DCT. Both transforms must be orthogonal to ensure invertibility. In this formulation, the transforms can be expressed as linear combinations of cosine basis functions:
\begin{equation}
    \boldsymbol{d}(k)=c(k)\sum\limits_{n=0}^{N-1}{\boldsymbol{l}(n)\cos \left( \frac{(2n+1)k\pi }{2N} \right)},
\end{equation}
\begin{equation}
    \boldsymbol{l}(n)=\sum\limits_{k=0}^{N-1}{c(k)\boldsymbol{d}(k)\cos \left( \frac{(2n+1)k\pi }{2N} \right)},
\end{equation}
\begin{equation*}
    \text{where $c(k)=$}
    \begin{cases}
        \sqrt{1/N}, &\text{if  $k = 0$.}\\
        \sqrt{2/N}, &\text{if  $k \ne 0$.}\\
    \end{cases}
\end{equation*}

Vital diagnostic information in normal ECG primarily lies in 0.5-50 Hz \cite{bib38}. Since the full spectrum of $\boldsymbol{d}$ extends beyond this range, we truncate $\boldsymbol{d}$ to retain only the informative components, which significantly reduces the computational burden and enhances feature learning efficiency. As shown in Fig.\autoref{fig2}, the high-frequency portion of $\boldsymbol{d}$ is near zero and contributes negligible useful information. Assuming a sampling frequency of $f_s$, the frequency resolution is $\Delta f=f_s / 2N$, and the frequency corresponding to the $k$-th coefficient $\boldsymbol{d}(k)$ is given by $f_k=k \cdot f_s/2N$. To preserve components below 50 Hz, we retain only the first $[50/\Delta f]$ coefficients. A series of validation experiments (Supplementary Note 2) demonstrates that signal fidelity is preserved despite the truncation.

\subsection{Discrete Time Conditional Diffusion}
We construct a diffusion model operating directly in the DCT domain, where both inputs and outputs are represented by truncated DCT coefficients. The \textit{forward diffusion process} $q$ and the \textit{reverse process} $p$ follow the canonical DDPM framework \cite{bib23} (Supplementary Note 3).

To align the reverse process $p$ with the true data distribution, it is standard practice to minimize the variational evidence lower bound (ELBO) of $-\log p(\boldsymbol{d}_0|\tilde{\boldsymbol{d}})$, where $\boldsymbol{d}_0$ denotes the clean DCT coefficient vector and $\tilde{\boldsymbol{d}}$ represents its noisy observation. As derived in \cite{bib39}, this objective simplifies to minimizing the distance between the true noise $\epsilon$ and its prediction from the noise predictor $\epsilon_{\theta}$. We employ a hybrid loss combining Charbonnier loss in both time and DCT domains \cite{bib33}. The overall loss function is represented as:
\begin{equation}
    \mathcal{L}_{hybrid}\left( \theta  \right)={{\mathbb{E}}_{{\boldsymbol{d}_0},\,\tilde{\boldsymbol{d}},\,t,\,\epsilon }}\left(\mathcal{L}_{DCT}+\mathcal{L}_{time}\right),
\end{equation}
with the DCT domain loss $\mathcal{L}_{DCT}$ and the time domain loss $\mathcal{L}_{time}$ defined as:
\begin{equation}
    \mathcal{L}_{DCT} = \sqrt{{\left\| \epsilon-{{\epsilon }_{\theta }}({\boldsymbol{d}_t},\,t,\tilde{\boldsymbol{d}}) \right\|}_2^2+\varepsilon}
\end{equation}
\begin{equation}
    \mathcal{L}_{time} = \sqrt{{\left\| \mathcal{D}^{-1}(\epsilon-{{\epsilon }_{\theta }}({{\boldsymbol{d}}_{t}},\,t,\tilde{\boldsymbol{d}})) \right\|}_2^2+\varepsilon}
\end{equation}
where $t$ denotes the timestep, $\boldsymbol{d}_t$ denotes the latent representation at step $t$ in the forward diffusion process $q$, $\mathcal{D}^{-1}(\cdot)$ denotes the 1D IDCT operation, and $\varepsilon$ is a small constant set to $10^{-4}$.

Given that diffusion sampling is inherently stochastic, simply averaging multiple reconstructions significantly reduces variance and improves fidelity \cite{bib18}. We therefore perform $m$ independent sampling runs and average the results. This ensemble strategy is referred to as the $m$-generation model or model-$m$.

\begin{figure*}[!t]
\centerline{\includegraphics[width=0.8\textwidth]{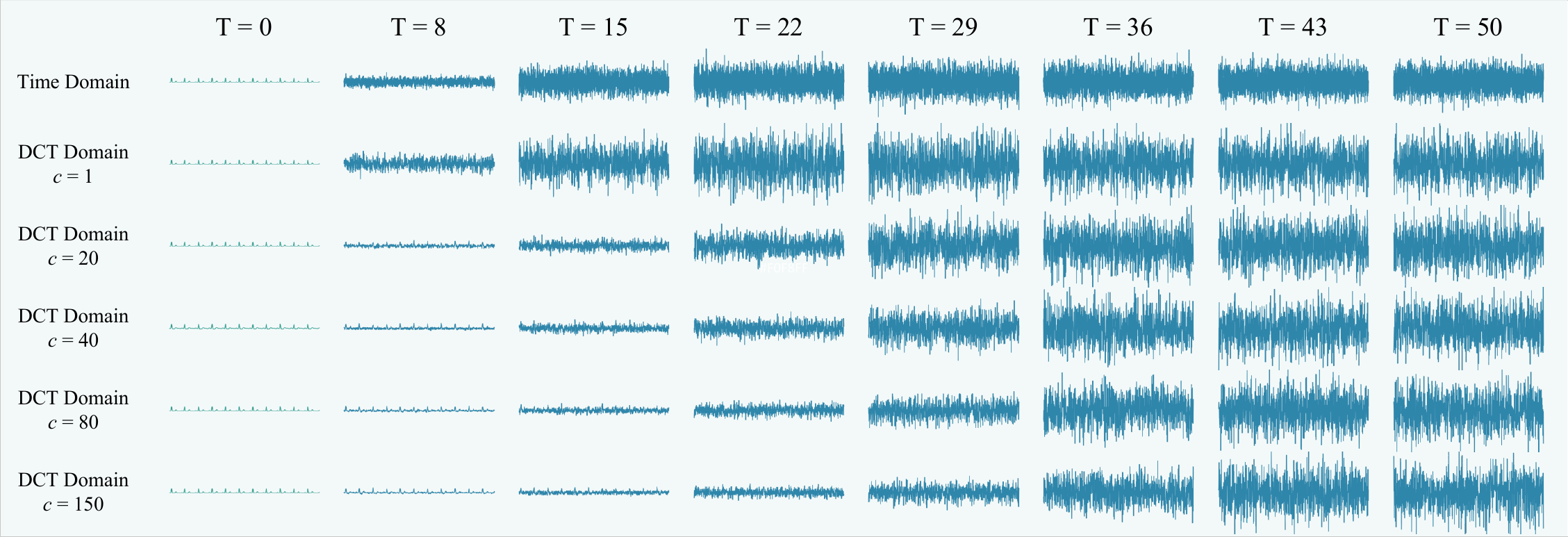}}
\caption{Forward diffusion in the time domain and DCT domain using the same quadratic noise schedule. (a) Visualization of the forward diffusion process under varying SNR scaling factors. (b) SNR as a function of timesteps.}
\label{fig3}
\end{figure*}

\begin{algorithm}[!t]
\caption{Training.}\label{alg1}
\begin{algorithmic}
\STATE \hspace{-0.5cm} \textbf{Input:} Distribution of signal pairs ${P_{data}}(\boldsymbol{l}_{clean},\boldsymbol{l}_{noisy})$, number of training iterations $N$, maximal diffusion timestep $T$, scaling factor for DCT coefficients $\eta$, hyperparameters for the noise schedule.
\STATE \hspace{-0.5cm} \textbf{Output:} Trained noise predictor $\epsilon_{\theta}$.
\end{algorithmic}
\begin{algorithmic}[1]
\STATE \textbf{for} $i=1,\dots,N$ \textbf{do}
\STATE \hspace{0.5cm}$({{\boldsymbol{l}}_{0}},\tilde{\boldsymbol{l}})\sim {{P}_{data}}({\boldsymbol{l}_{clean}},{\boldsymbol{l}_{noisy}})$
\STATE \hspace{0.5cm}$\boldsymbol{d}_0 = \mathcal{D}_{trunc}(\boldsymbol{l}_0) / \eta$
\STATE \hspace{0.5cm}$\tilde{\boldsymbol{d}} = \mathcal{D}_{trunc}(\tilde{\boldsymbol{l}}) / \eta$
\STATE \hspace{0.5cm}$\epsilon \sim \mathcal{N}(0,\mathbf{I})$
\STATE \hspace{0.5cm}$t\sim \text{Uniform}(\{1,\dots,T\})$
\STATE \hspace{0.5cm}\text{Take gradient descent step on}
\STATE \hspace{1cm}${{\nabla }_{\theta }}[\mathcal{L}_{hybrid}\left( \theta  \right)]$
\STATE \textbf{end for}
\end{algorithmic}
\end{algorithm}

\subsection{DCT Coefficients Scaling}
\label{subsec:DCT Coefficients Scaling}
It is recognized that inputs to the diffusion model must be normalized to a bounded range $[-1,1]$ to ensure training stability. However, the distribution of DCT coefficients is highly skewed and spans several orders of magnitude. We observe that naively scaling all DCT coefficients by a global bound leads to poor training dynamics. Inspired by the Entropy-Consistent Scaling approach \cite{bib30}, we decompose the DCT coefficients into a direct current (DC) component and an alternating current (AC) component. The AC coefficients are typically concentrated near zero but exhibit a much larger range due to occasional extreme values. To enable stable diffusion, we scale all DCT coefficients using the bound derived solely from the DC component.

To estimate this bound robustly, we construct a dataset of 32578 noisy ECG signals (detailed in Section\autoref{subsec:Preprocessing}), which are representative of the inputs encountered during diffusion, and convert them into truncated DCT coefficients. From the DC components of this dataset, we compute the Monte Carlo estimation of the scaling bound $\eta$, using percentile-based truncation to mitigate the impact of outliers:
\begin{equation}
    \eta =\max \left( \left| {{P}_{\tau }} \right|,\left| {{P}_{100-\tau }} \right| \right),
\end{equation}
where $P_{\tau}$ denotes the $\tau\text{-th}$ percentile of the DC component distribution.

We visualize the symmetric percentile $P_{\tau}$ and $P_{100-\tau}$ for various values of $\tau$ (Fig. S2). Based on empirical validation across multiple trials, we select $\tau$ = 1.75, which yields an $\eta$ value of approximately 3. Before injecting into the diffusion model, we scale both the clean DCT coefficient vector $\boldsymbol{d}_0$ and its noisy observation $\tilde{\boldsymbol{d}}$ by default, namely $\boldsymbol{d}_0 \gets \boldsymbol{d}_0 / \eta$, $\tilde{\boldsymbol{d}} \gets \tilde{\boldsymbol{d}}/\eta$.

\begin{algorithm}[!t]
\caption{Sampling.}\label{alg2}
\begin{algorithmic}
\STATE \hspace{-0.5cm} \textbf{Input:} trained noise predictor $\epsilon_{\theta}$, noisy signal $\tilde{\boldsymbol{l}}$, maximal diffusion timestep $T$, number of generations $m$, hyperparameters for the noise schedule.
\STATE \hspace{-0.5cm} \textbf{Output:} Denoised signal $\hat{\boldsymbol{l}}$
\end{algorithmic}
\begin{algorithmic}[1]
\STATE $\boldsymbol{d}_0^0=0$
\STATE $\tilde{\boldsymbol{d}} = \mathcal{D}_{trunc}(\tilde{\boldsymbol{l}}) / \eta$
\STATE \textbf{for} $i=1,\dots,m$ \textbf{do}
\STATE \hspace{0.5cm} $\boldsymbol{d}_T \sim \mathcal{N}(0,\mathbf{I})$ 
\STATE \hspace{0.5cm} \textbf{for} $t=T,\dots,1$ \textbf{do}
\STATE \hspace{1cm}$z \sim \mathcal{N}(0,\mathbf{I})$ \textbf{if} $t>1,$ \textbf{else} $z=0$
\STATE \hspace{1cm}${\boldsymbol{d}_{t-1}}={{\mu }_{\theta }}({\boldsymbol{d}_{t}},t,\tilde{\boldsymbol{d}})+{{\sigma }_{\theta }}({\boldsymbol{d}_{t}},t,\tilde{\boldsymbol{d}})\cdot z$
\STATE \hspace{0.5cm} \textbf{end for}
\STATE \hspace{0.5cm} $\boldsymbol{d}_0^i=\boldsymbol{d}_0^{i-1}+\boldsymbol{d}_0$
\STATE \textbf{end for}
\STATE $\hat{\boldsymbol{d}}=\boldsymbol{d}_0^m/m$
\STATE $\hat{\boldsymbol{l}}=\mathcal{D}^{-1}_{pad}(\eta\hat{\boldsymbol{d}})$
\STATE return $\hat{\boldsymbol{l}}$
\end{algorithmic}
\end{algorithm}

\begin{figure*}[!t]
\centerline{\includegraphics[width=0.8\textwidth]{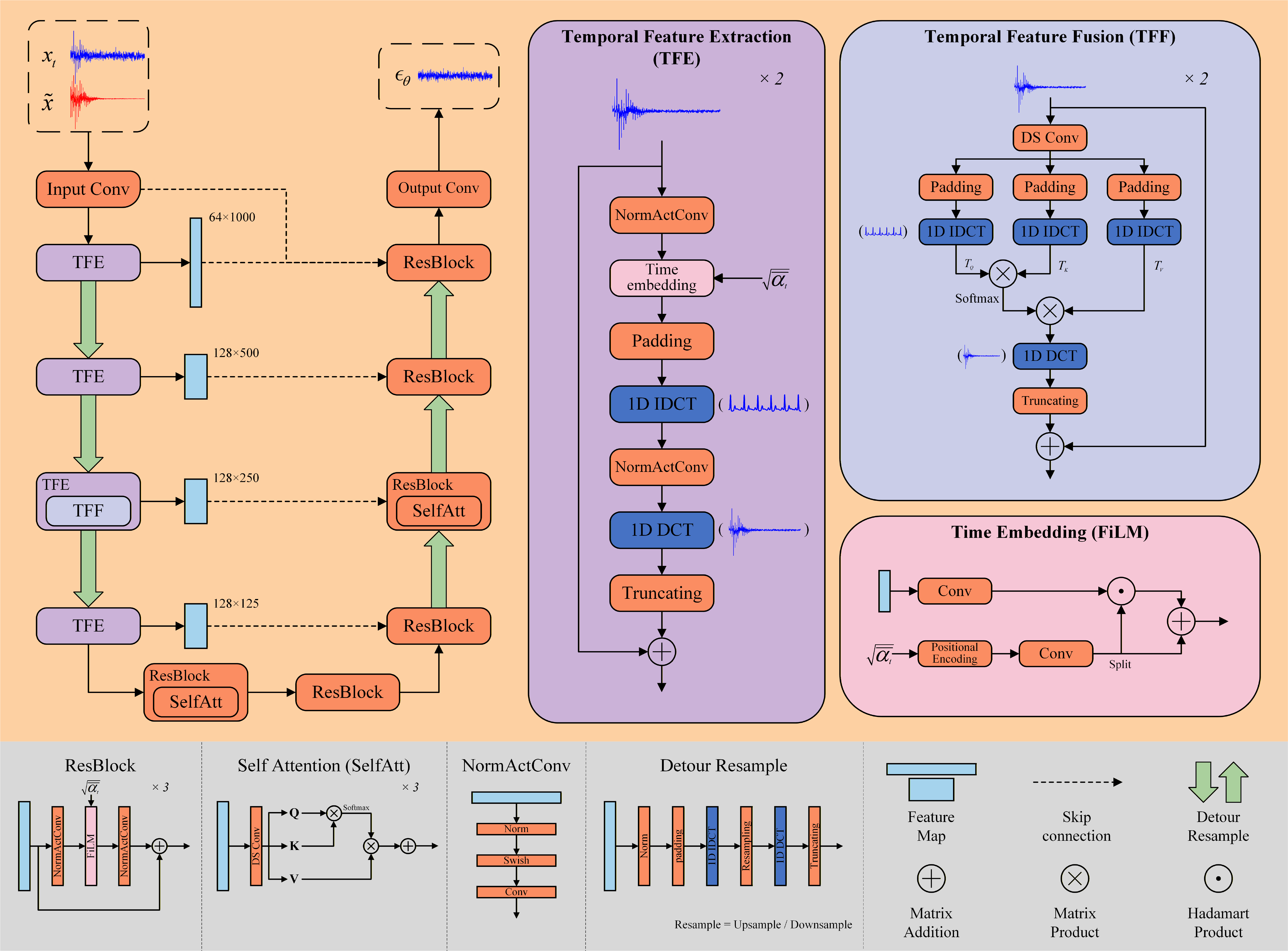}}
\caption{The architecture of the noise predictor, which comprises an U-Net backbone, TFE, TFF and other constitutive modules.}
\label{fig4}
\end{figure*}

\subsection{Noise Schedule and SNR Scaling}
\label{subsec:Noise Schedule and SNR Scaling}
The noise schedule defines how noise is introduced as a function of timesteps during the diffusion process, and ensures that, by the final timestep, the data distribution is fully transformed into an isotropic Gaussian. A widely adopted choice is the variance-preserving noise schedule, which guarantees that the variance of any latent representation $\boldsymbol{d}_t$ remains bounded throughout the diffusion process. We initially define a quadratic variance-preserving noise schedule based on a total of $T=50$ timesteps and boundary values $\beta_1=10^{-4}$, $\beta_T=0.5$:
\begin{equation}
    {{\beta }_{t}}={{\left( \sqrt{{{\beta }_{1}}}+\left( t-1 \right)\frac{\sqrt{{{\beta }_{T}}}-\sqrt{{{\beta }_{1}}}}{T-1} \right)}^{2}},
\end{equation}
\begin{equation}\label{equation8}
    {{l}_{t}}=1-{{\beta }_{t}},\quad{{\bar{l}}_{t}}=\prod\limits_{s=1}^{t}{{{l}_{t}}}, \quad t\in\{1,\dots,T\}.
\end{equation}

Given this schedule, the signal-to-noise ratio (SNR) of the latent representation $\boldsymbol{d}_t$ at timestep $t$ is computed as:
\begin{equation}\label{equation9}
    \text{SNR}(t)=\frac{{{{\bar{l}}}_{t}}}{1-{{{\bar{l}}}_{t}}}.
\end{equation}

In contrast to the time domain, a notable property of DCT is that most energy is concentrated in low-frequency components, while high-frequency coefficients tend toward zero. Consequently, during the forward diffusion process, high-frequency components are rapidly overwhelmed by noise, making it difficult for the noise predictor to learn fine-grained denoising behavior. Moreover, our global scaling of all DCT coefficients further intensifies this imbalance, suggesting that standard noise schedules designed for time-domain data are not suitable for the DCT domain.

To address this issue, we introduce an SNR scaling factor $c$ to modulate the noise progression in diffusion. The modified SNR is defined as $\text{SNR}'(t)=c\cdot \text{SNR}(t)$. Substituting into \eqref{equation8}, we derive the parameters ${\bar{\gamma }}_{t}$ analogous to $\bar{l}_{t}$ for the new noise schedule:
\begin{equation}
    {{\bar{\gamma }}_{t}}=\frac{\text{SNR}'(t)}{1+\text{SNR}'(t)}.
\end{equation}

Subsequently, referring to \eqref{equation9}, we can iteratively solve the per-timestep parameters corresponding to $\beta_t$, $l_t$ to construct the complete scaled noise schedule.

Fig.\autoref{fig3} visualizes the forward diffusion process under different scaling factors $c$, with all data converted back to the time domain for intuitive comparison. As $c$ increases, the rate of noise addition slows, preserving ECG structure longer during diffusion. We empirically find that the diffusion model yields the best performance when $c$ is set to 150. At this value, frequency-domain corruption remains milder than in the time-domain counterpart for timestep $t<35$, allowing the model to better learn the noise distribution.

Additionally, we adopt the hierarchical uniform sampling approach proposed in \cite{bib40}, which enables the noise predictor to generalize across continuous noise levels. Specifically, we first construct the sequence $S=\{1,\sqrt{{{{\bar{\gamma }}}_{1}}},\dots,\sqrt{{{{\bar{\gamma }}}_{T}}}\}$, and then sample $\sqrt{{{{\bar{\alpha }}}_{t}}}\sim \text{Uniform}({{S}_{t-1}},{{S}_{t}})$, where $\bar{\alpha}_t$ corresponds to $\bar{l}_t$ in the original formulation.

With the DCT-domain representation, normalization strategy, and noise schedule now established, the fundamental workflow of TFCDiff is in place. To ensure clarity and reproducibility, the comprehensive training and sampling procedures are summarized in Algorithm\autoref{alg1} and Algorithm\autoref{alg2}, respectively. Specifically, we define the operator $\mathcal{D}_{trunc}(\cdot)$ as the process of applying 1D DCT to a signal of length $N$ and retaining only the first $N'$ low-frequency coefficients. Conversely, the operator $\mathcal{D}^{-1}_{pad}(\cdot)$ reconstructs the signal by zero-padding the truncated coefficients from $N'$ back to $N$, followed by 1D IDCT. For $\mu_\theta$ and $\sigma_\theta$ please refer to Supplementary Note 3.

\subsection{Network Architecture of the Noise Predictor}
The architecture of the noise predictor, as depicted in Fig.\autoref{fig4}, incorporates an encoder-decoder framework developed from the classical U-Net \cite{bib41}. The encoder learns global contextual features by progressively downsampling the extracted representations, while the decoder reconstructs the compressed features to the original dimension through successive upsampling. Skip connections bridge corresponding encoder and decoder layers at multiple scales to recover information lost during downsampling. To further enhance the model's capacity to discern complex inter-channel dependencies, we integrate Squeeze-and-Excitation (SE) blocks into these skip connections \cite{bib49}, \cite{bib50}. This mechanism adaptively recalibrates channel-wise feature responses based on global context, enabling the network to dynamically amplify informative components while suppressing noise-corrupted channels.

The fundamental building block of the network is a residual block comprising Group Normalization, Swish activation, and 1D convolution. To enhance the capacity for capturing long-range dependencies, self-attention modules are integrated into the middle decoder layer and the bottleneck layer.

Since the inputs consist of DCT coefficients, which may lead to loss of temporal details, we propose a Temporal Feature Enhancement Mechanism (TFEM) to mitigate this limitation. At each scale of the encoder, we construct a stream that alternates between frequency-domain and time-domain features, inspired by \cite{bib31}, \cite{bib33}, to achieve comprehensive cross-domain integration. We also investigate applying TFEM in the decoder, but this yields no performance improvement while increasing computational cost.

TFEM is composed of two modules: Temporal Feature Extraction (TFE) and Temporal Feature Fusion (TFF). In TFE, to convert a feature map from the DCT domain to the time domain, zero padding is first applied to restore its original length, followed by 1D IDCT. After processing through a residual block, the feature map is transformed back into the frequency domain by 1D DCT and subsequent truncation. At the middle encoder layer, we further fuse these heterogeneous representations from two inconsistent domains using TFF. Depthwise separable convolution is used to encode the feature map $F$ into $Q=W_{d}^{Q}W_{p}^{Q}F$, $K=W_{d}^{K}W_{p}^{K}F$, $V=W_{d}^{V}W_{p}^{V}F$. Zero padding and 1D IDCT are then applied to generate time-domain tensors $T_Q$, $T_K$ and $T_V$. Long-range correlations and local similarities in the time domain are jointly modeled via matrix multiplication:
\begin{equation}
    {F}_{fused}={{T}_{V}}\times \text{Softmax}\left( \frac{{{T}_{Q}}\times {{T}_{K}^\text{T}}}{\sqrt{C}} \right),
\end{equation}
where $C$ denotes the number of channels in the input feature map.

Because the main feature stream remains in the frequency domain, direct upsampling or downsampling induces aliasing artifacts that impair training stability. To alleviate the problem, we introduce a detour resample strategy. The feature map is first converted to the time domain, then upsampled or downsampled using strided convolutions or interpolation, and finally transformed back into the DCT domain.

For conditional injection of the noisy observation $\tilde{\boldsymbol{d}}$, we adopt the approach from \cite{bib42}. $\tilde{\boldsymbol{d}}$ is concatenated with the latent representation $\boldsymbol{d}_t$ at the network input and jointly embedded via a convolutional layer. Regarding time embedding, we employ feature-wise linear modulation (FiLM) with affine transformation \cite{bib43}. The noise level $\sqrt{\bar{\alpha}_t}$ is first encoded by sinusoidal positional embeddings \cite{bib44}, then passes through a linear layer to produce a scale vector $\gamma$ and a shift vector $\beta$. The modulation operation on an input feature map $F_\text{in}$ is given by:
\begin{equation}
    {F}_{out}=(1+\gamma )\odot {{F}_{in}}+\beta.
\end{equation}

\begin{figure*}[!t]
\centerline{\includegraphics[width=0.9\textwidth]{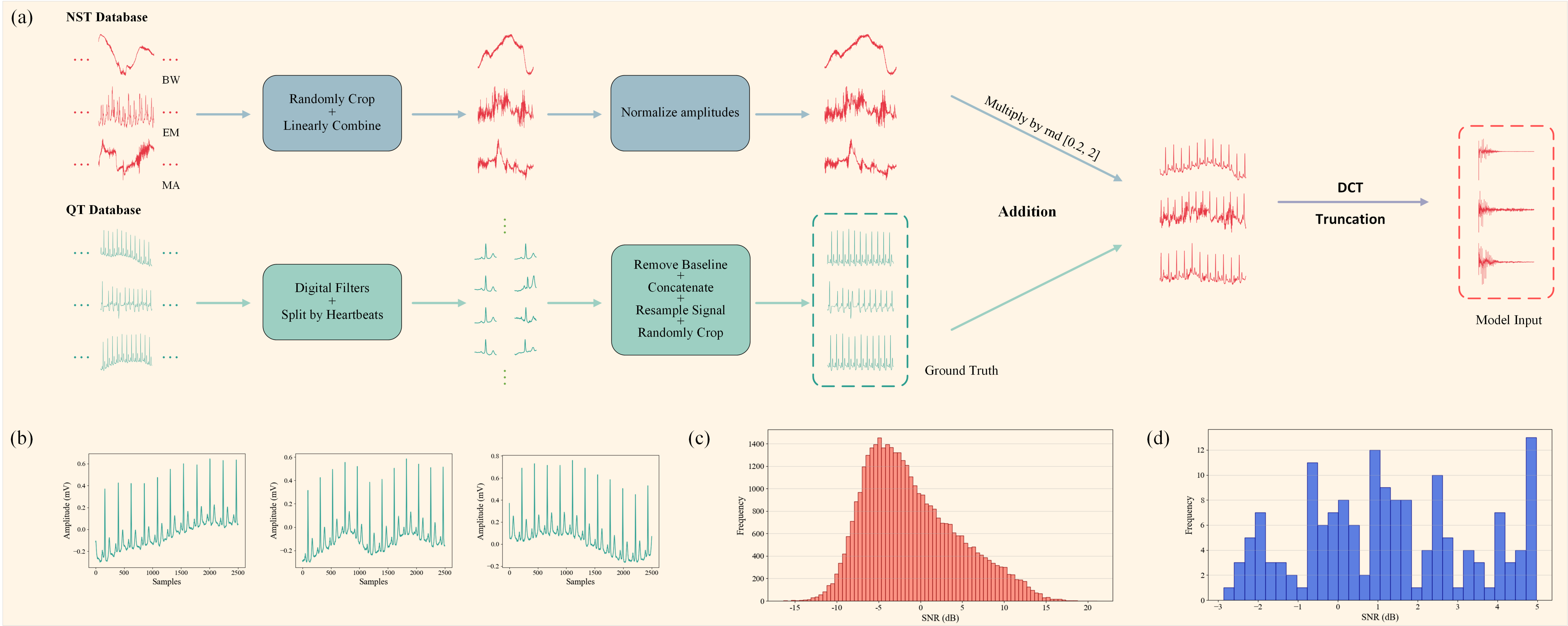}}
\caption{Dataset preprocessing. (a) An overview. (b) Three 10-second sequences sampled from the QT Database. (c) SNR distribution of the synthesized dataset. (d) SNR distribution of SimEMG Database.}
\label{fig5}
\hypertarget{fig5:sub}{}
\end{figure*}

\section{Experiments}
\subsection{Datasets}
Two datasets are utilized to train and evaluate the performance of TFCDiff, including the QT Database \cite{bib45} and the MIT-BIH Noise Stress Test (MIT-BIH NST) Database \cite{bib46}, both available at www.physionet.org \cite{bib47}. To assess the generalization of the model, we additionally introduce an external public dataset named the SimEMG database \cite{bib21}. Summary of all three datasets are listed in Table S1.

\subsubsection{QT Database} The QT database comprises high-quality dual-channel ECG Holter recordings aggregated from seven distinct sources, providing ground truth $\boldsymbol{l}_0$ for training and testing. It features a wide variety of QRS and ST-T morphologies, reflecting diverse cardiac physiology. Each recording includes annotations of P-QRS-T waveform boundaries.

\subsubsection{MIT-BIH NST Database} From the NST database, we extract noise excerpts to contaminate clean ECG signals $\boldsymbol{l}_0$, thereby generating noisy observations $\tilde{\boldsymbol{l}}$. Electrodes are placed on the thighs and arms, with lead axes configured to cancel out ECG components, yielding pure noise recordings. These include the three primary types of noise found in ambulatory ECG signals: baseline wander (BW), muscle artifact (MA) also known as electromyographic (EMG) noise, and electrode motion artifact (EM). 

\subsubsection{SimEMG Database} The SimEMG Database provides the first collection of authentic paired EMG-contaminated and EMG-free ECG signals acquired from 15 healthy volunteers. The voltage is sampled at a resolution of 200 points per millivolt. EMG-contaminated signals are recorded via electrodes placed on the hands, while clean reference signals are obtained from electrodes on the shoulders. This data differs fundamentally from synthesized signals and is therefore used for inter-dataset testing to evaluate the generalization of TFCDiff.

\subsection{Preprocessing}
\label{subsec:Preprocessing}
Previous studies typically focused on single-beat ECG segments and omitted filtering preprocessing. However, in multi-beat ECG signals, such as those shown in Fig. \hyperlink{fig5:sub}{5(b)}, baseline wander becomes non-negligible, rendering raw QT Database signals unsuitable as direct ground truth. To address this, we first apply digital filters including a bandpass filter and a median filter to remove impulsive artifacts. Inspired by \cite{bib48}, we then propose a piecewise fitting method for baseline wander removal that preserves ECG morphology while leveraging the annotations of the waveform boundaries in the QT Database. Specifically, we use annotations of QRS complexes to segment each ECG recording. Within each segment, we assume the baseline wander can be approximated by a linear function. After subtracting the fitted baseline, we concatenate the corrected segments and apply Hermite interpolation at the junctions to ensure smooth transitions, resulting in clean ECG signals.

After baseline wander removal, we allocate 91 ECG recordings from the QT Database to the training set and the remaining 14 to the test set, consistent with prior work \cite{bib11}, \cite{bib18}, \cite{bib19}. The recording IDs of the test set are listed in Supplementary Table S2, with two records selected from each of the seven source datasets in the QT Database to better validate model generalization. Simultaneously, we partition the NST Database by channel and time: the first half of Channel 1 is assigned to the training set, and the second half of Channel 2 is reserved for testing, thereby preventing data leakage.

For noise addition, the QT Database signals are first oversampled to 360 Hz to align with the NST Database. A clean segment $\boldsymbol{l}_0$ with a length of 3600 is then randomly cropped from it. Separately, three noise segments $e_1$, $e_2$, $e_3$, corresponding to BW, MA and EM, are extracted from the NST Database, each matching the length of $\boldsymbol{l}_0$. We employ a fRMN strategy, where the composite noise RMN is defined as $e=re_1+me_2+ne_3$, with $r,m,n\ge0$ and $r+m+n=1$. The resulting noise $e$ is normalized to align its amplitude with that of $\boldsymbol{l}_0$, scaled by an intensity factor $\lambda \in [0.2,2]$ to modulate the noise level, and added to $\boldsymbol{l}_0$ as follows:
\begin{equation}
    \tilde{\boldsymbol{l}}={\boldsymbol{l}_{0}}+\lambda \frac{\max ({\boldsymbol{l}_{0}})-\min ({\boldsymbol{l}_{0}})}{\max (e)-\min (e)}e
\end{equation}

As a result, we yield 37590 synthesized data pairs, of which 32578 are used for training and 5012 for testing. Fig. \hyperlink{fig5:sub}{5(c)} illustrates the SNR distribution of this dataset. The complete preprocessing pipeline is summarized in Fig. \hyperlink{fig5:sub}{5(a)}.

For the SimEMG Database, the signals are undersampled to 360 Hz and uniformly segmented into 10-second intervals. We only retain those segments with SNR below 5 dB as \cite{bib19}, resulting in 158 data pairs for inter-dataset testing. The SNR distribution for SimEMG is shown in Fig. \hyperlink{fig5:sub}{5(d)}. To convert the raw SimEMG digital counts into physical voltage units (mV) consistent with the QT Database, the signals were divided by the specified gain factor of 200 samples/mV.

\begin{table*}[!t]
\caption{Overall Comparison Results of Different Methods for ECG Denoising on the Synthesized Dataset. The Noise Level Ranges from 0.2 to 2.}
\label{table1}
\setlength{\tabcolsep}{0cm}
\centering
\begin{tabular}{
    >{\centering\arraybackslash}p{2.20cm}
    >{\centering\arraybackslash}p{2.20cm}
    >{\centering\arraybackslash}p{2.20cm}
    >{\centering\arraybackslash}p{2.20cm}
    >{\centering\arraybackslash}p{2.20cm}
    >{\centering\arraybackslash}p{2.20cm}
    }
\toprule[1.3pt]
Models & SSD ($\mu$V$^2$) $\downarrow$ & MAD (mV) $\downarrow$ & PRD (\%) $\downarrow$ & CosSim $\uparrow$ & ImSNR (dB) $\uparrow$ \\ 
\midrule[0.8pt]
CBAM-DAE & 226.144±496.562 & 0.981±0.613 & 2047.762±756.884 & 0.252±0.097 & 1.078±5.590 \\
FIR & 114.049±293.828 & 0.597±0.586 & 50.357±19.293 & 0.834±0.124 & 6.390±3.438 \\
IIR & 117.897±294.303 & 0.647±0.630 & 51.864±18.747 & 0.826±0.124 & 6.038±3.617 \\
DRNN & 69.887±161.657 & 0.601±0.563 & 69.366±27.043 & 0.839±0.104 & 7.048±4.578 \\
FCN-DAE & 38.279±69.373 & 0.533±0.469 & 50.623±28.772 & 0.890±0.097 & 8.917±5.129 \\
DeepFilter & 37.008±66.841 & 0.392±0.335 & 51.382±21.874 & 0.897±0.070 & 8.937±4.617 \\
DesCod-1 & 34.002±87.379 & 1.107±1.078 & 39.986±21.688 & 0.910±0.097 & 10.636±5.122 \\
DesCod-3 & 25.064±68.798 & 0.756±0.604 & 34.630±20.361 & 0.939±0.069 & 12.274±4.873 \\
DesCod-5 & 23.636±67.801 & 0.647±0.500 & 33.148±20.300 & 0.946±0.061 & 12.776±4.775 \\
DesCod-10 & 22.312±64.709 & 0.526±0.430 & 31.755±20.102 & 0.952±0.055 & 13.274±4.691 \\
TCDAE & 20.985±51.394 & 0.322±0.334 & 29.612±13.892 & 0.952±0.051 & 12.669±4.727 \\
\midrule[0.8pt]
TFCDiff-1 & 27.649±83.464 & 0.351±0.430 & 31.702±21.450 & 0.945±0.075 & 12.974±4.284 \\
TFCDiff-3 & 21.500±69.620 & 0.310±0.396 & 27.790±20.404 & 0.960±0.057 & 14.446±4.354 \\
TFCDiff-5 & 20.279±66.075 & 0.302±0.385 & 26.876±20.105 & 0.963±0.052 & 14.818±4.377 \\
TFCDiff-10 & \textbf{19.399±63.326} & \textbf{0.295±0.377} & \textbf{26.139±20.104} & \textbf{0.966±0.049} & \textbf{15.135±4.425} \\
\bottomrule[1.3pt]
\end{tabular}
\end{table*}

\begin{figure*}[!t]
\centerline{\includegraphics[width=0.8\textwidth]{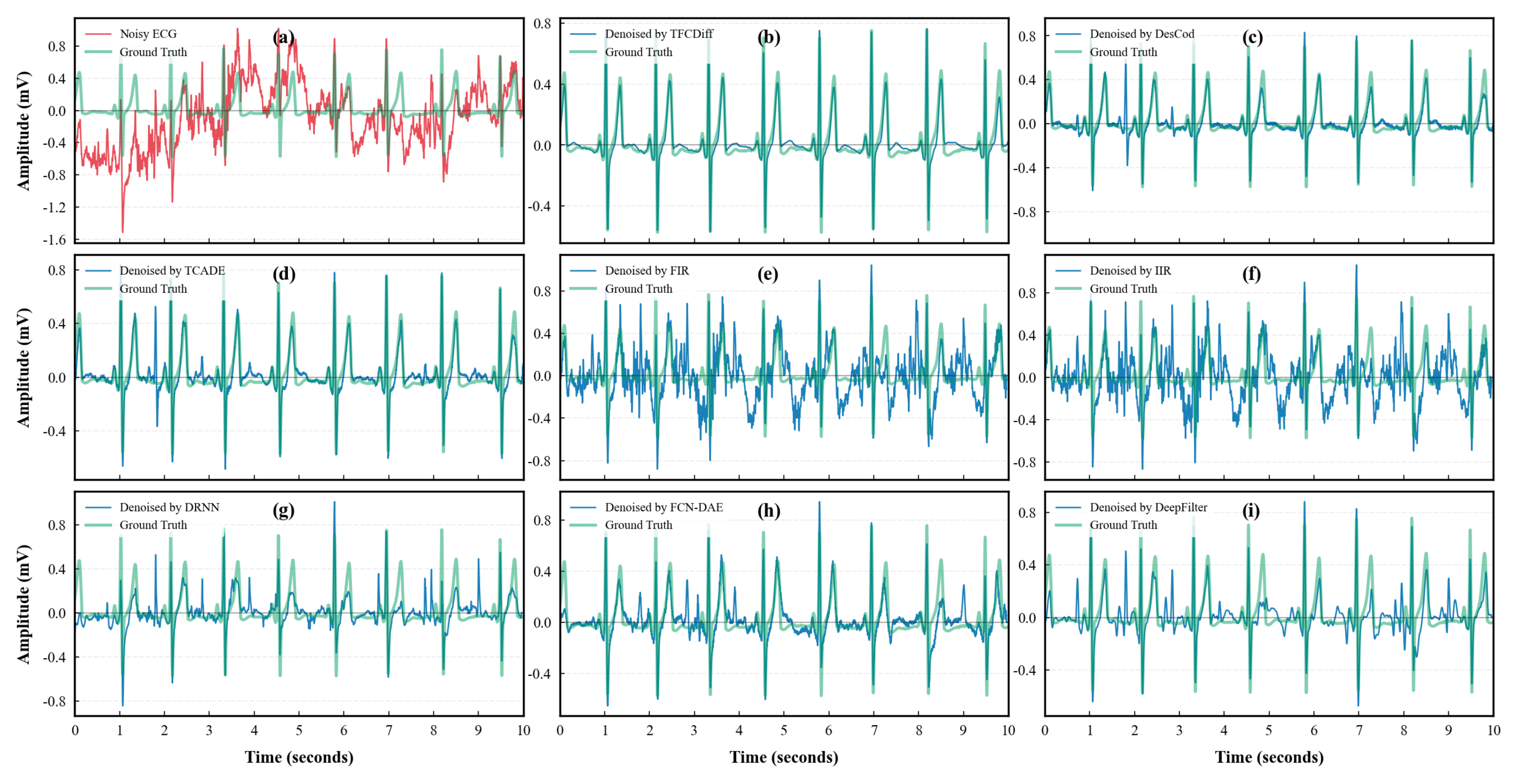}}
\caption{Visual comparison of denoising results between different methods on a 10-second ECG segment from the synthesized dataset. The red line in (a) is the noisy ECG, the green lines in (a) -- (i) represent the ground truth signal, and the blue lines in (b) -- (i) depict denoised signals by different methods.}
\label{fig6}
\end{figure*}

\subsection{Evaluation Metrics}
To evaluate denoising performance, we employ five quantitative metrics widely used in prior research \cite{bib11}, \cite{bib18}, \cite{bib19}, measuring distortion relative to the ground truth. Sum of the Square of the Distances (SSD), Absolute Maximum Distance (MAD) and Percentage Root-Mean-Square Difference (PRD) are distance-based metrics, while Cosine Similarity (CosSim) assesses similarity by computing the cosine of the angle between two vectors. Improved SNR (ImSNR) quantifies the enhancement in signal quality achieved through denoising. The definitions of metrics are detailed in Supplementary Note 4. The lower the SSD, MAD and PRD, and the higher the CosSim and ImSNR, the better the denoising result.

\begin{figure*}[!t]
\centerline{\includegraphics[width=0.8\textwidth]{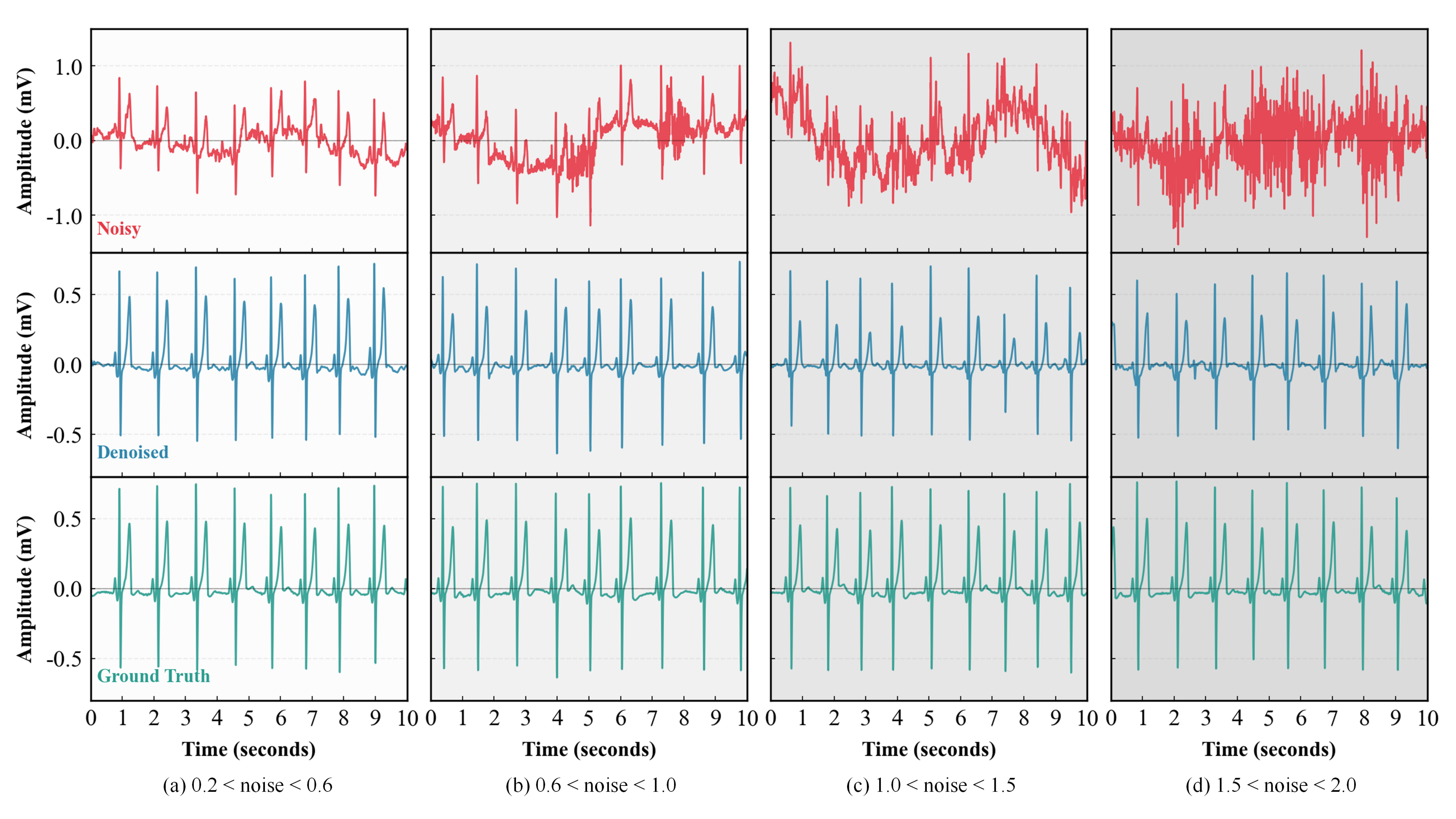}}
\caption{Visualization of TFCDiff denoised results on the synthesized dataset with representative samples. (a) Noise Level 0.2-0.6: SSD 2.156 $\mu$V$^2$, MAD 0.138 mV, PRD 15.162 \%, CosSim 0.989, ImSNR 15.204 dB. (b) Noise Level 0.6–1.0: SSD 5.200 $\mu$V$^2$, MAD 0.150 mV, PRD 24.140 \%, CosSim 0.979, ImSNR 16.266 dB. (c) Noise Level 1.0–1.5: SSD 16.416 $\mu$V$^2$, MAD 0.333 mV, PRD 53.270 \%, CosSim 0.945, ImSNR 14.866 dB. (d) Noise Level 1.5–2.0: SSD 12.989 $\mu$V$^2$, MAD 0.289 mV, PRD 42.170 \%, CosSim 0.944, ImSNR 13.442 dB.}
\label{fig7}
\end{figure*}

\subsection{Comparative Methods and Implementation Details}
We conduct comparative experiments against eight benchmark methods, grouped into three categories: digital filters, end-to-end deep learning filters, and diffusion-based filters. Digital filters include Finite Impulse Response (FIR) and Infinite Impulse Response (IIR) filters, which contain no learnable parameters. End-to-end deep learning filters encompass diverse architectures such as CNN, LSTM and Transformer, specifically: FCN-DAE \cite{bib13}, DRNN \cite{bib10}, DeepFilter \cite{bib11}, CBAM-DAE \cite{bib14}, and TCDAE \cite{bib16}. DeScoD is the only diffusion-based method. Both DeScoD and TCDAE reported SOTA performance for ECG denoising in 2024. All referenced methods are implemented using their officially released code, with the input length set to 3600 to evaluate performance on multi-beat signals. Exceptions are FCN-DAE and CBAM-DAE, whose input length is adjusted to 3584 to avoid shape-mismatch errors caused by multiple downsampling. All methods use identical data preparation, namely, all trainable models are trained on the same training set, whereas all methods including non-trainable ones are evaluated on the same test set to ensure fair comparison.

For our proposed method, we implement the model using the PyTorch framework. Training employs the Adam optimizer with an initial learning rate of $10^{-3}$, reduced by a factor of 0.1 every 150 epochs. The total number of epochs is 400, with a batch size of 128. During training, 30\% of the training data is held out exclusively for validation, and the model weights yielding the best validation performance are selected for final evaluation on the test set.

\begin{table*}[!t]
\caption{Overall Comparison Results of Different Methods for ECG Denoising on the SimEMG Database.}
\label{table2}
\setlength{\tabcolsep}{0cm}
\centering
\begin{tabular}{
    >{\centering\arraybackslash}p{2.20cm}
    >{\centering\arraybackslash}p{2.20cm}
    >{\centering\arraybackslash}p{2.20cm}
    >{\centering\arraybackslash}p{2.20cm}
    >{\centering\arraybackslash}p{2.20cm}
    >{\centering\arraybackslash}p{2.20cm}
    }
\toprule[1.3pt]
Models & SSD ($\mu$V$^2$) $\downarrow$ & MAD (mV) $\downarrow$ & PRD (\%) $\downarrow$ & CosSim $\uparrow$ & ImSNR (dB) $\uparrow$ \\ 
\midrule[0.8pt]
FIR & 23.280±9.703 & 0.396±0.096 & 59.844±8.810 & 0.803±0.066 & -0.044±0.153\\
IIR & 26.445±13.728 & 0.448±0.178 & 61.650±8.875 & 0.789±0.071 & -0.484±0.825\\
DRNN & 12.203±5.647 & 0.420±0.102 & 50.924±6.418 & 0.870±0.033 & 2.653±1.845\\
FCN-DAE & 8.503±4.584 & 0.310±0.128 & 45.904±6.522 & 0.887±0.035 & 4.263±2.283\\
DeepFilter & 7.594±2.590 & 0.212±0.057 & 56.580±10.326 & 0.855±0.052 & 4.633±0.928\\
DesCod-1 & 5.640±7.796 & 1.015±1.071 & 34.439±12.546 & 0.930±0.065 & 7.381±3.636\\
DesCod-3 & 3.066±2.773 & 0.617±0.572 & 28.487±7.782 & 0.957±0.034 & 9.057±2.629\\
DesCod-5 & 2.568±1.444 & 0.504±0.428 & 26.848±6.029 & 0.963±0.019 & 9.575±2.473\\
DesCod-10 & 2.172±0.832 & 0.388±0.300 & 25.220±4.741 & 0.968±0.013 & 10.117±2.167\\
TCDAE & 3.790±1.281 & 0.181±0.042 & 29.112±3.133 & 0.970±0.008 & 7.653±1.664\\
\midrule[0.8pt]
TFCDiff-1 & 2.428±1.265 & 0.150±0.047 & 25.454±4.253 & 0.969±0.013 & 9.773±1.952\\
TFCDiff-3 & 2.004±1.068 & 0.141±0.046 & 23.214±3.838 & 0.975±0.010 & 10.627±1.943\\
TFCDiff-5 & 1.925±1.093 & 0.139±0.047 & 22.717±3.864 & 0.976±0.010 & 10.834±1.995\\
TFCDiff-10 &\textbf{1.849±1.015} & \textbf{0.137±0.045} & \textbf{22.344±3.784} & \textbf{0.977±0.010} & \textbf{10.986±1.983}\\
\bottomrule[1.3pt]
\end{tabular}
\end{table*}

\section{Results}
\subsection{Intra-dataset Testing}
On the synthesized dataset, we compare TFCDiff with eight benchmark methods and report evaluation metrics as mean values and standard deviations in Table\autoref{table1}. TFCDiff-10 demonstrates exceptional performance with values of 19.399 ± 63.326 $\mu$V$^2$, 0.295 ± 0.377 mV, 26.139 ± 20.104 \%, 0.966 ± 0.049 and 15.135 ± 4.425 dB for SSD, MAD, PRD, CosSim and ImSNR respectively, establishing new benchmarks for ECG denoising across all five metrics. Notable failures occur in CBAM-DAE, which exhibits catastrophic performance degradation, hence it is excluded form subsequent comparisons. Due to the inherent stochasticity of diffusion models, denoising performance consistently improves as the count of generations used for averaging increases. Remarkably, TFCDiff-1 outperforms most benchmark models, including the refined DesCod-10 in MAD. For the strongest competitor TCDAE, TFCDiff-3 achieves performance comparable to it, while TFCDiff-5 comprehensively surpasses it across all metrics. To further visualize the preservation of critical morphological features, we randomly select a 10-second ECG segment to illustrate the denoising results of various methods in Fig.\autoref{fig6}. Notably, comparative methods suffer from severe morphological failures, characterized by the hallucination of false peaks or the persistence of pervasive noise. In contrast, TFCDiff preserves the complete morphological structure of the ECG signal without introducing such artifacts.

To further assess model robustness, we conduct a comparative study under different noise levels defined by the intensity factor. Results are systematically presented in Supplementary Table S3 - S7, each dedicated to one evaluation metric. TFCDiff maintains consistent superiority across most metrics at all noise levels, and demonstrates the highest stability as noise intensity increases. It remains effective even under the most challenging 1.5 - 2.0 interval where noise energy exceeds that of the original ECG components. 

Visual evidence in Fig.\autoref{fig7} illustrates denoised time-domain waveforms under four noise intervals. TFCDiff accurately preserves critical ECG morphologies, and achieves robust ECG signal reconstruction despite drastic noise interference. Complementary time-frequency analysis via Short-Time Fourier Transform (Fig. S3) further validates the effective suppression of noise components.

\begin{table*}[!t]
\caption{Ablation Study on the Impact of DCT Domain Diffusion and TFEM.}
\label{table3}
\setlength{\tabcolsep}{0cm}
\centering
\begin{tabular}{
    >{\centering\arraybackslash}p{1.00cm}
    >{\centering\arraybackslash}p{1.00cm}
    >{\centering\arraybackslash}p{1.00cm}
    >{\centering\arraybackslash}p{2.20cm}
    >{\centering\arraybackslash}p{2.20cm}
    >{\centering\arraybackslash}p{2.20cm}
    >{\centering\arraybackslash}p{2.20cm}
    >{\centering\arraybackslash}p{2.20cm}
    }
\toprule[1.3pt]
TD & DCT & TFEM & SSD ($\mu$V$^2$) $\downarrow$ & MAD (mV) $\downarrow$ & PRD (\%) $\downarrow$ & CosSim $\uparrow$ & ImSNR (dB) $\uparrow$ \\ 
\midrule[0.8pt]
\checkmark & -- & -- & 64.131±139.266 & 0.549±0.565 & 60.874±30.432 & 0.832±0.134 & 7.370±3.830\\
\checkmark & -- & \checkmark & 65.608±350.499 & 0.448±0.628 & 44.650±35.089 & 0.909±0.109 & 10.590±5.049\\
-- & \checkmark & -- & 98.477±214.740 & 0.495±0.557 & 74.927±62.565 & 0.791±0.1849 & 7.262±4.126\\
-- & \checkmark & \checkmark & \textbf{27.649±83.464} & \textbf{0.351±0.430} & \textbf{31.702±21.450} & \textbf{0.945±0.075} & \textbf{12.974±4.284}\\
\bottomrule[1.3pt]
\multicolumn{8}{p{13.2cm}}{\vspace{-0.1cm}\hspace{0.3cm}TD refers to time-domain diffusion, and DCT refers to DCT diffusion. All results are from a single generation.}
\end{tabular}
\end{table*}

\begin{figure*}[!t]
\centerline{\includegraphics[width=0.8\textwidth]{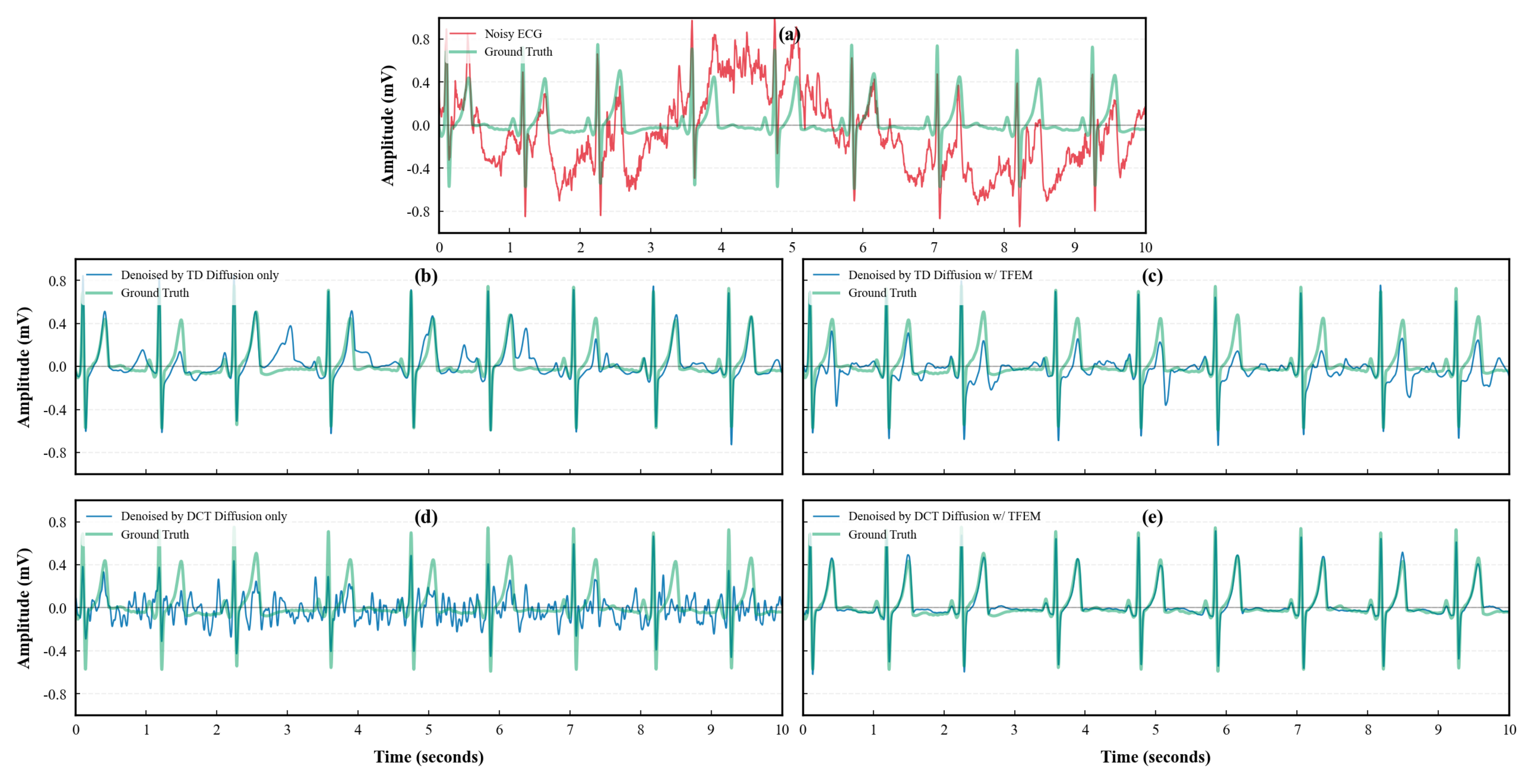}}
\caption{Visual comparison of the denoised results in the ablation study of DCT diffusion and TFEM. The red line in (a) is the noisy ECG, the green lines in (a) -- (e) represent the ground truth signal, and the blue lines in (b) -- (e) depict denoised signals under different configurations: (b) Time-domain diffusion only, (c) Time-domain diffusion with TFEM, (d) DCT diffusion only, (e) DCT diffusion with TFEM.}
\label{fig8}
\hypertarget{fig8:sub}{}
\end{figure*}

\subsection{Inter-dataset Testing}
To evaluate cross-dataset generalization, we conduct comparative experiments on the SimEMG Database, as documented in Table\autoref{table2}. It should be emphasized that SimEMG contains real EMG-contaminated ECG signals, fundamentally distinct from the synthesized dataset used for training. TFCDiff achieves the lowest reconstruction error, consistent with its intra-dataset performance, and outperforms its closest competitor, TCDAE, across nearly all metrics even with only one generation. This persistent advantage highlights the capacity to learn essential noise distributions, enabling precise separation of ECG signals from unseen noise patterns. 

Notably, DesCod, another diffusion model that underperforms on the synthesized dataset, surpasses other end-to-end methods including TCDAE. This implies the intrinsic generalization potential of diffusion models for physiological signal restoration. 

Fig. S4 visualizes the superior denoising performance of TFCDiff on EMG-contaminated ECG signals. Complementary time-frequency analysis (Fig. S5) further corroborates the effective suppression of EMG artifacts. These results demonstrate the efficacy and robustness of our approach in denoising multi-beat ECG signals, leveraging DCT diffusion coupled with TFEM.
 
\begin{table*}[!t]
\caption{Ablation Study on the Impact of SE Block and Hybrid Loss.}
\label{table4}
\setlength{\tabcolsep}{0cm}
\centering
\begin{tabular}{
    >{\centering\arraybackslash}p{1.50cm}
    >{\centering\arraybackslash}p{1.50cm}
    >{\centering\arraybackslash}p{2.20cm}
    >{\centering\arraybackslash}p{2.20cm}
    >{\centering\arraybackslash}p{2.20cm}
    >{\centering\arraybackslash}p{2.20cm}
    >{\centering\arraybackslash}p{2.20cm}
    }
\toprule[1.3pt]
SE Block& Hybrid Loss & SSD ($\mu$V$^2$) $\downarrow$ & MAD (mV) $\downarrow$ & PRD (\%) $\downarrow$ & CosSim $\uparrow$ & ImSNR (dB) $\uparrow$ \\ 
\midrule[0.8pt]
--  & -- & 41.800±187.452 & 0.368±0.415 & 40.546±22.759 & 0.912±0.097 & 10.563±4.153\\
\checkmark & -- & 36.602±119.564 & 0.360±0.438 & 36.148±24.383 & 0.927±0.094 & 11.671±4.287\\
-- & \checkmark & 28.444±95.862 & 0.357±0.442 & \textbf{30.482±23.695} & \textbf{0.947±0.081} & \textbf{13.476±4.354}\\
\checkmark & \checkmark & \textbf{27.649±83.464} & \textbf{0.351±0.430} & 31.702±21.450 & 0.945±0.075 & 12.974±4.284\\
\bottomrule[1.3pt]
\end{tabular}
\end{table*}

\section{Discussion}
\subsection{Ablation Study}
\label{subsec:Time-Frequency Complementary Mechanism}
\subsubsection{Time-Frequency Complementary Mechanism}
TFCDiff incorporates two key designs to bridge time and DCT domains, i.e., a diffusion model operating in the DCT domain and a TFEM-integrated U-Net for noise prediction. To evaluate their impact, we conduct an ablation study. Starting from a baseline time-domain (TD) diffusion model with a plain U-Net architecture, we incrementally incorporate the DCT diffusion workflow and TFEM, resulting in four configurations: time-domain diffusion only, time-domain diffusion with TFEM, DCT diffusion only, and DCT diffusion with TFEM. Note that the time-domain diffusion operates on temporal signals, so TFEM within it reverses the order of forward and inverse DCT to enhance spectral features instead. This minor modification is still consistent with the fundamental design of TFEM. 

As shown in Table\autoref{table3}, introducing TFEM significantly improves the performance of the time-domain diffusion, demonstrating its universal utility for denoising. Interestingly, the standalone DCT diffusion model exhibits substantial performance degradation, but augmenting it with TFEM triggers a dramatic reversal, surpassing both time-domain diffusion variants. Further insights emerge from Fig.\autoref{fig8}, which visualizes denoised ECG signals to showcase waveform details. Fig. \hyperlink{fig8:sub}{8(b)} and \hyperlink{fig8:sub}{(c)} reveal that the baseline time-domain diffusion model introduces false waveforms that distort physiological information, an issue partially corrected by adding TFEM. By contrast, Fig. \hyperlink{fig8:sub}{8(d)} shows that the standalone DCT diffusion model struggles to address pervasive artifacts, which significantly degrades overall signal integrity. Fig. \hyperlink{fig8:sub}{8(e)} illustrates the efficacy of TFEM, which resolves the challenge of artifact suppression and elevates reconstruction quality to a new level. Complementary time-frequency analysis via STFT (Fig. S6) delves deeper, providing spectral evidence of how noise patterns are selectively suppressed while physiological components remain intact. These observations suggest that DCT diffusion can excel at ECG feature extraction but may struggle to suppress pervasive noise. TFEM acts as a mediator, dynamically integrating temporal and spectral representations to refine details. The collaborative implementation of DCT diffusion and TFEM thus delivers optimal outcomes by leveraging their time-frequency complementary strengths.

\subsubsection{SE Block and Hybrid Loss}
To assess the effectiveness of the SE block and the hybrid loss function within TFCDiff, we conduct ablation experiments on the synthesized dataset. All models are evaluated under a single-generation inference setting. We compare four configurations: (a) Baseline, devoid of both SE blocks and hybrid loss. (b) SE-only, incorporating SE blocks into the U-Net skip connections. (c) Loss-only, utilizing the hybrid Charbonnier loss in both time and DCT domains without SE blocks. (d) Full TFCDiff, integrating both components.

Table\autoref{table4} summarizes the quantitative performance of these variants. The results indicate that introducing either the SE block or the hybrid loss individually yields significant improvements over the baseline, reinforcing the importance of adaptive feature extraction and multi-domain context modeling for efficient ECG signal processing. Interestingly, while the Loss-only variant achieves superior mean scores on several metrics compared to the Full TFCDiff, the full configuration demonstrates lower variance across trials. We attribute this to a regularization trade-off, as SE-induced feature gating likely constrains the hybrid loss optimization to prevent overfitting, thereby sacrificing marginal mean gains for enhanced stability. Since clinical applications prioritize consistent reliability, we selected the full configuration for its robust quality and minimized variance.

\begin{table}[!t]
\caption{Comparison of Inference Latency (ms) on a 10-second ECG recording for different methods.}
\label{table5}
\setlength{\tabcolsep}{0cm}
\centering
\begin{tabular}{
    >{\centering\arraybackslash}p{1.25cm}
    >{\centering\arraybackslash}p{1.25cm}
    >{\centering\arraybackslash}p{1.25cm}
>{\centering\arraybackslash}p{1.25cm}
>{\centering\arraybackslash}p{1.25cm}
>{\centering\arraybackslash}p{1.25cm}
>{\centering\arraybackslash}p{1.25cm}
    }
\toprule[1.3pt]
Models & FCN-DAE & DRNN & DeepFilter & TCDAE & DesCod-1 & TFCDiff-1\\ 
\midrule[0.8pt]
Time (ms) & 19.09 & 7.56 & 47.57 & 54.30 & 36.21 & 43.72\\
\bottomrule[1.3pt]
\end{tabular}
\end{table}

\subsection{Inference Latency}
As detailed in Table\autoref{table5}, we evaluate the inference latency of TFCDiff and comparative deep learning methods on a 10-second ECG recording using a standard desktop configuration with an Intel\textregistered\ Core\texttrademark\ i7-13700K CPU (3.40 GHz) and an NVIDIA GeForce RTX 4090 D GPU. All reported latencies fall within the thresholds required for real-time denoising requirements. Notably, leveraging just-in-time (JIT) compilation significantly accelerates inference, likely by optimizing the computational graph of the DCT operations. While Table\autoref{table5} reports the latency for a single generation, a multi-generation ensemble naturally increases computational cost. However, in practical deployment, this overhead is effectively alleviated using overlapping sliding windows that average predictions within redundant regions, and the generation speed can be further enhanced through batch computation. Although TFCDiff exhibits slightly higher latency than some baselines, its superior denoising fidelity justifies this cost.

\begin{table*}[!t]
\caption{Fiducial Point Detection Performance on the Synthesized Dataset between different methods.}
\label{table6}
\centering
\setlength{\tabcolsep}{0pt} 

\begin{tabular}{
    >{\centering\arraybackslash}p{2.20cm}
    >{\centering\arraybackslash}p{1.30cm}
    >{\centering\arraybackslash}p{1.30cm}
    >{\centering\arraybackslash}p{1.30cm}
    >{\centering\arraybackslash}p{1.30cm}
    >{\centering\arraybackslash}p{1.30cm}
    >{\centering\arraybackslash}p{1.30cm}
    >{\centering\arraybackslash}p{1.30cm}
    >{\centering\arraybackslash}p{1.30cm}
    >{\centering\arraybackslash}p{1.30cm}
    >{\centering\arraybackslash}p{1.30cm}
    }
    
\toprule[1.3pt]
\multirow{2}{*}{Models} & 
\multicolumn{2}{c}{P-wave} & 
\multicolumn{2}{c}{Q-wave} & 
\multicolumn{2}{c}{R-wave} & 
\multicolumn{2}{c}{S-wave} & 
\multicolumn{2}{c}{T-wave} \\ 

\cmidrule(lr){2-3} 
\cmidrule(lr){4-5} 
\cmidrule(lr){6-7} 
\cmidrule(lr){8-9} 
\cmidrule(lr){10-11}

& F1 (\%) & MAE (ms) & F1 (\%) & MAE (ms) & F1 (\%) & MAE (ms) & F1 (\%) & MAE (ms) & F1 (\%) & MAE (ms) \\

\midrule[0.8pt]
Raw Data & 88.64 & 84.90 & 91.17 & 39.15 & 91.08 & 29.18 & 91.07 & 33.10 & 69.71 & 288.51 \\
DeepFilter & 86.27 & 117.83 & 93.05 & 27.65 & 92.98 & 19.51 & 93.08 & 21.60 & 64.93 & 346.90 \\
DesCod-10 & 93.43 & \textbf{61.05} & 97.27 & 17.37 & 97.23 & 12.44 & 97.27 & 15.36 & 83.11 & 171.56 \\
TCADE & 93.68 & 64.19 & 97.71 & 14.26 & 97.57 & 9.04 & 97.61 & 11.25 & 84.27 & 159.16 \\
TFCDiff-10 & \textbf{94.65} & 61.63 & \textbf{98.55} & \textbf{13.39} & \textbf{98.56} & \textbf{7.30} & \textbf{98.55} & \textbf{9.31} & \textbf{90.24} & \textbf{104.35} \\
\bottomrule[1.3pt]
\end{tabular}
\end{table*}

\begin{figure*}[!t]
\centerline{\includegraphics[width=0.8\textwidth]{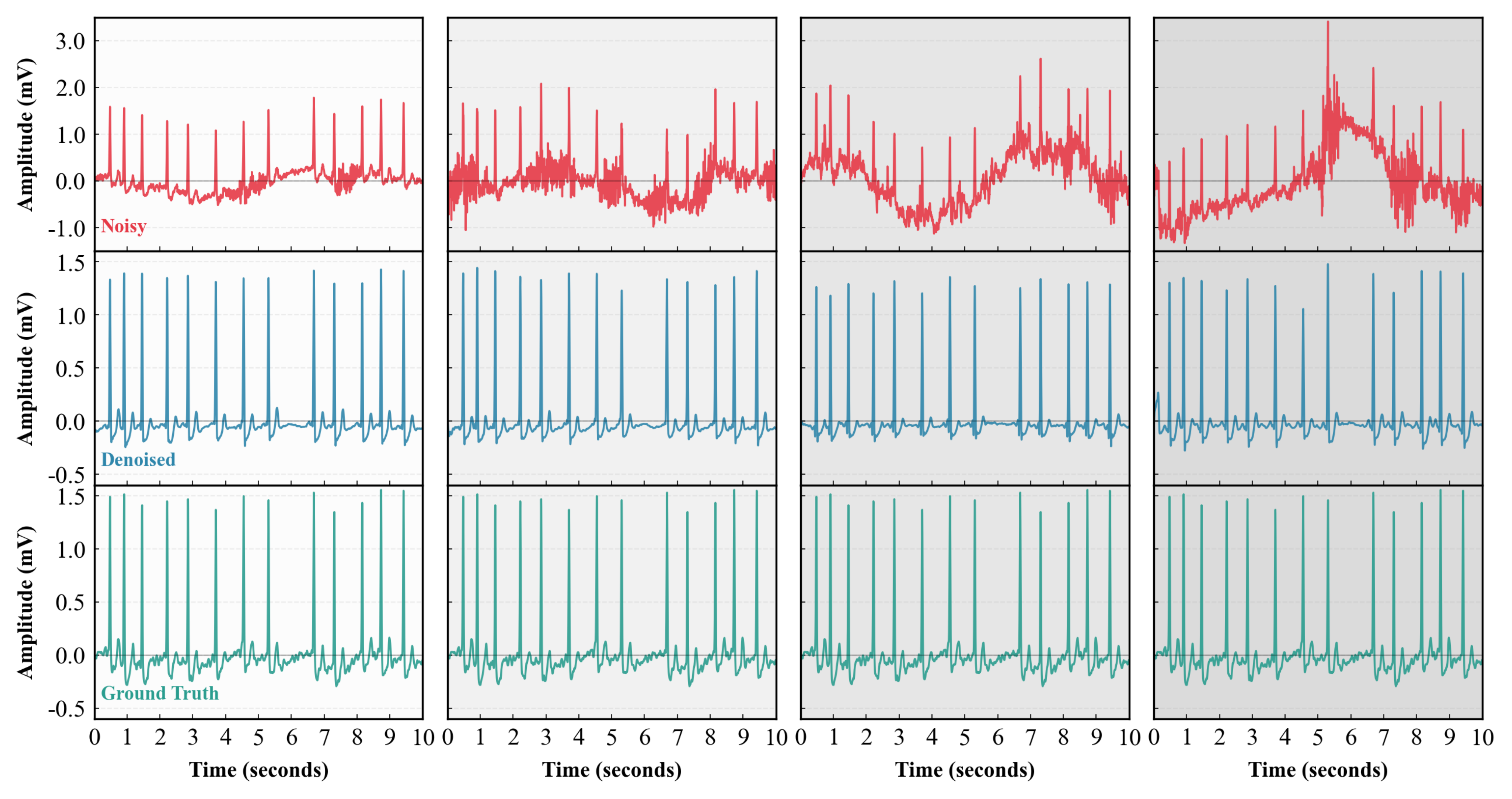}}
\caption{Visualization of TFCDiff denoised results on AF signals across varying noise intensities.}
\label{fig9}
\end{figure*}

\subsection{Practical ECG Denoising}
\subsubsection{Preservation of Diagnostic Morphology}
High global metrics do not guarantee clinical utility if critical morphological features such as P waves or ST segments undergo excessive smoothing or shifting. To validate morphological fidelity, we detect five distinct fiducial points (P, Q, R, S, and T waves) across clean, noisy, and processed signals within the synthesized dataset using a peak-prominence-based algorithm implemented by NeuroKit2 \cite{bib51}, \cite{bib52}. Treating detections from clean signals as ground truth allows us to compute F1-score (F1) and Mean Absolute Error (MAE) for each fiducial point (Supplementary Note 5). We only compare the performance of TFCDiff with DeepFilter, DesCod-10 and TCDAE, which perform well on global metrics.

As shown in Table\autoref{table6}, TFCDiff-10 outperforms these competitors across nearly all morphological indicators by delivering superior detection accuracy and temporal precision. Notably, TFCDiff achieves substantial improvements in T-Peak reconstruction which is essential for identifying ischemia and predicting sudden cardiac death. These findings confirm that TFCDiff effectively reconstructs diagnostic details while preserving physiological integrity to ensure the clinical reliability of denoised signals.

\subsubsection{Robustness to ECG Anomalies}
In practical ECG denoising, maintaining fidelity across diverse cardiac rhythms is paramount for clinical reliability. We therefore evaluate TFCDiff on pathological anomalies, specifically Atrial Fibrillation (AF) and Premature Ventricular Contractions (PVCs) obtained from \cite{bib53}. Following the protocol in Section IV-B, we construct synthesized data pairs spanning four noise intensities. For AF signals, where P-waves are absent, the model effectively suppresses noise while preserving the characteristic irregular baseline as shown in Fig.\autoref{fig9}. Although extreme noise intensities induce distortions in fine pathological features such as fibrillatory waves, the global diagnostic structure remains intact. Similar robustness extends to PVC cases (Fig. S7). These results confirm that TFCDiff generalizes well beyond normal sinus rhythms, ensuring applicability in real-world scenarios involving complex arrhythmia.

\subsubsection{Processing of Multi-Beat Segments}
Previous studies on ECG denoising predominantly focus on short sequences containing only a single heartbeat \cite{bib10}, \cite{bib11}, \cite{bib14}, \cite{bib17}, \cite{bib18}, \cite{bib19}, a constraint that imposes significant clinical limitations. First, real-world ECG recordings often lack precise beat annotations, making accurate waveform segmentation particularly challenging under heavy noise conditions. Second, even when heartbeats are successfully segmented, padding these fragments to meet fixed model input requirements is inevitable. The subsequent concatenation process to restore the original signal length frequently distorts critical heartbeat intervals, especially R-R intervals, which are essential for arrhythmia diagnosis. By contrast, our proposed TFCDiff model is trained directly on raw 10-second ECG signals. This end-to-end approach eliminates the need for complex preprocessing and segmentation, thereby preserving temporal integrity and facilitating seamless deployment in real-world scenarios.

\subsubsection{Adaptability to Dynamic Noise}
Furthermore, most existing models are trained exclusively on fixed noise patterns \cite{bib11}, \cite{bib12}, \cite{bib17}, \cite{bib18}, \cite{bib19}, \cite{bib20}, limiting their generalization capability. In practice, however, noise sources are highly diverse and dynamic: variations in respiratory rate during physical activity induce Baseline Wander (BW), muscular contractions generate Muscle Artifact (MA), and body movements coupled with sweat-induced impedance changes affect Electrode Motion (EM). Consequently, models trained on single noise types or uniformly mixed distributions often fail to adapt to the complex noise profiles encountered in clinical measurements. To address this, we employ the fRMN strategy to mimic these realistic, diverse noise combinations consisting of BW, MA, and EM. This training paradigm ensures that TFCDiff remains robust even during strenuous physical activity, enabling reliable continuous surveillance and early detection of cardiac abnormalities within flexible wearable monitoring scenarios.

\subsection{Limitations and Future Works}
There are several limitations in our study which need to be addressed in future work: 1) The generation speed of TFCDiff remains to be optimized, primarily due to its iterative sampling constraints. As accelerated sampling via DPM solver shows promise \cite{bib26}, a subsequent direction will involve extending our diffusion framework to a continuous-time form and reformulating our sampling process as a deterministic ODE to apply DPM solver. In addition, we plan to simultaneously apply structural pruning to the U-Net backbone of the noise predictor and implement model quantization to minimize latency and memory footprint for resource-constrained devices. 2) We observe unsatisfactory denoising results for signals with intensive EMG noise. This is likely attributable to the similar distribution between EMG noise and ECG signals. Efforts could be directed toward exploring flow-matching techniques \cite{bib27}, which offer greater flexibility for modeling arbitrary distribution. 3) The effectiveness of TFCDiff on intelligent ECG monitoring devices remains unproven. Building upon our developed wearable ECG monitoring prototype (Fig. S8), we plan to deploy TFCDiff on this platform to evaluate its practicality in real-world scenarios.

\section{Conclusion}
In this paper, we propose TFCDiff, a generative diffusion model for denoising ECG signals contaminated by complex mixed noise. By modeling directly in the frequency domain, TFCDiff efficiently reconstructs periodic waveforms such as P-QRS-T complexes while reducing computational overhead. To further integrate time-frequency information, we design a U-Net architecture leveraging TFEM, wherein TFE and TFF modules collaboratively preserve critical physiological waveforms and refine signal details. TFCDiff trains and performs inference on 10-second sequences, with the fRMN strategy enhancing its capacity to learn diverse noise patterns. This enables plug-and-play downstream applications without preprocessing. Compared to eight ECG denoising benchmarks, TFCDiff achieves SOTA performance, particularly demonstrating significant superiority in inter-dataset generalization tests. Its robustness under extreme noise conditions extends the monitoring scenario of wearable ECG devices to high-intensity motion backgrounds.

\end{document}